\documentstyle[a4,twoside]{article}
\newcommand{\f}{\mbox{\Eufb f}}
\newcommand{\g}{\mbox{\Eufb g}}

\newcommand{\PP}{\mbox{\bf P}}
\newcommand{\QQ}{\mbox{\bf Q}}
\newcommand{\ap}{\mbox{$a^{+}$}}
\newcommand{\am}{\mbox{$a^{-}$}}

\newcommand{\e}[1]{\epsilon^{#1}}

\newfont{\Eufm}{eufm10}
\newcommand{\C}{\mbox{\Eufm C}}
\newcommand{\R}{\mbox{\Eufm R}}

\newfont{\Eufb}{eufb10}
\newcommand{\JJ}{\mbox{\Eufb J}}
\newcommand{\HH}{\mbox{\underline{\bf H}}}
\newcommand{\KK}{\mbox{\underline{\bf K}}}

\newcommand{\J}{\makebox[1ex][c]{%
\makebox[-0.3ex][l]{\rule[0.6ex]{0.8ex}{0.4pt}}%
\makebox[0ex][l]{\rm J}              }}

\makeatletter
\renewcommand{\@oddhead}{Among Quadratic Hamiltonians \hfill \thepage}
\renewcommand{\@evenhead}{\thepage \hfill S. A. Choro\v{s}avin }
\renewcommand{\@oddfoot}{}
\renewcommand{\@evenfoot}{}
\makeatother

{\vspace*{2ex}\par {#1}\quad \it }{\medskip\par }

\newenvironment{Thm}[2]{\par\addvspace{\bigskipamount}\noindent%
{\bf #1#2}\it \hspace*{10ex}}%
{\par\addvspace{\bigskipamount} }

\newenvironment{Definition}[1]{\begin{Thm}{Definition}{#1}\rm }{\end{Thm} }

\newenvironment{Theorem}[1]{\begin{Thm}{Theorem}{#1}}{\end{Thm} }

\newenvironment{Proposition}[1]{\begin{Thm}{Proposition}{#1}}{\end{Thm} }

\newenvironment{Corollary}[1]{\begin{Thm}{Corollary}{#1}}{\end{Thm} }
\newenvironment{Observation}[1]{\begin{Thm}{Observation}{#1}}{\end{Thm} }
\newenvironment{Remark}[1]{\begin{Thm}{Remark}{#1}\rm }{\end{Thm} }

\newenvironment{Example}[1]{\begin{Thm}{Example}{#1}}{\end{Thm} }

\newenvironment{Proof}{\par\addvspace{\bigskipamount} {\sc Proof}}%
{\par\hspace*{\fill}$\Box$ \par\addvspace{\bigskipamount} }

\title{Among Quadratic Hamiltonians, Bogoliubov Transformations
       and Non-Regular States on CCRs *-Algebra. \protect\\
       I. Pure and Invariant States. \\
   (In the Mood for the Manuceau Verbeure Theorems \protect\\
 about Quasi-free States and Automorphisma of the CCR Algebra) 
            }
\author{ S. A. Choro\v{s}avin }
\begin{document}
\maketitle

\begin{abstract} 
 The paper's features are these: 

 1) we discuss {\bf especially } quadratic (alias bilinear) 
   Bose-Hamiltonians, the related Bogoliubov transformations 
   and {\bf especially } quasi-free-like (alias coherent or Fock-like) 
   states 

 2) we discuss {\bf any } quadratic Bose-Hamiltonians and 
    Bogoliubov transformations, 
    whether diagonalizable or not,
    whether proper or improper,  
    and {\bf arbitrary } quasi-free-like states, 
    whether regular or non-regular they are 

 3) we associate notions and terms of the CCRs 
\footnote{ CCRs = Canonical Commutation Relations } 
 theory 
 with notions and terms of the indefinite inner product spaces theory. 
 Then, we apply the corresponding `bilingual dictionary' 
 so as to construct invariant states of some of the quadratic Hamiltonians.

\end{abstract}

    \newpage

\section{Introduction}

 Quadratic (bilinear) {\bf Fermi}-Hamiltonians 
 have a very attractive property. 
 One can diagonalize them and the way is {\bf not unique}, 
 as a result, given a Fermi-Hamiltonian, there exists 
 {\bf a Fock-like invariant} state, and such a state is {\bf not unique}.  
 \footnote{ Fock-like state = even quasi-free state 
 }. 
 Nothing really like this is present in the case of quadratic 
 {\bf Bose}-Hamiltonians.
 By the contrast, there are cases, 
 e.g., the case of the repulsive oscillator, 
 where one cannot diagonalize the Bose-Hamiltonian and
 Fock-like invariant states do not exist at all. 

 Here we shall say more precisely: 
 there exists no Fock-like invariant state 
 with {\bf continuous} characteristic function.  

 But why `to diagonalize' and even why `{\bf continuous}'? 
 One needs firstly {\bf invariant} states with a definite 
 algebraic structure. 

 We have tried to compose suitable constructions 
 and here, in this paper, they are presented. 
\footnote{
 To be more precise, I must say: this paper is 
 a very large abstract of 
 [Ch1], [Ch2], [Ch3] (and the summarizing [Ch4]), 
 but proofs of theorems omitted. 
 }

 \newpage   
 \section{ Prerequisites: The Quadratic Hamiltonians,
   Canonical Commutation Relations, 
   and Bogoliubov Transformations
 }
\footnote{
  Throughout this paper we assume 
 that units of measure   
 are choosen and fixed 
 so that 
$\hbar = 1$ and so that 
 the momentum quantity, 
{\PP}, 
 and the position quantity, 
\QQ, 
 both become dimensionless quantities. 
 } 

 This section consists primarily of 
 formal constructions and manipulations
 as we 
 would like
 briefly to explain
 what we mean by 
%
 `{\bf Quadratic Hamiltonians, CCRs = Canonical Commutation Relations}'
 and `{\bf Bogoliubov (Canonical)Transformations}.'
 
 The Quadratic Hamiltonian of an $N$-degree of freedom system
\footnote{ $N$ may be infinite } 
 is here a formal expression
$$
 h={\sum}_{k,l}
 \left(s_{k\,l}a_k^*a_l-\frac{1}{2}\overline{t_{k\,l}}a_k a_l
                     -\frac{1}{2}t_{k\,l}a_l^*a_k^* \right)
$$
 where 
$$
    s_{k\,l}=\overline{s_{l\,k}}\;;\;t_{l\,k}=t_{k\,l}
$$
 and 
$ a_k^*, a_l ; k,l=1,\ldots, N $
 are thought of 
 as elements of an (associative) *-algebra.

 We will suppose 
 $ a_k^*, a_l ; k,l=1,\ldots, N $
 to be subject to the relations
$$
 [a_k,a_l^*]=(a_ka_l^*-a_l^*a_k)=\delta_{k\,l}\;;
 [a_k,a_l]=0\;;[a_l^*,a_k^*]=0
$$
  The relations are said to be 
 the {\bf Canonical Commutations Relations in the Fock-Dirac form}.  

 Let us write 
$$
 u:=(u_1,...,u_N)\;;\;
 u^+:=(u^+_1,...,u^+_N)\;;\;
 u^-:=(u^-_1,...,u^-_N)\;;\;
$$
\begin{eqnarray*}
 \ap(u)&:=& u_1 a_1^*+\cdots u_N a_N^*  \\
 a(u)  &:=& (\ap(u)^{*}  \equiv \overline{u_1}a_1+\cdots \overline{u_N} a_N \\ 
 \am(u)&:=& a(\overline u)  \equiv {u_1}a_1+\cdots {u_N} a_N 
                            \equiv \ap(\overline u)^* 
\end{eqnarray*}
$
 \mbox{ ( hence } \ap(u) =  a(u)^*  = \am(\overline u)^*   \mbox{ ) }
$ 
  and in addition we set 
$$
 A(u^+\oplus u^-):= \ap(u^+) + \am(u^-)
 \equiv  u^+_1 a_1^*+\cdots u^+_N a_N^*+u^-_1 a_1+\cdots u^-_N a_N
$$

 Formal calculations show that 
$$
 [h,A(u^+\oplus u^-)]=A({u^+}'\oplus {u^-}')\,,
$$
 where 
${u^+}'\oplus {u^-}'$ 
 is defined by
$$
 {{u^+}'\choose {u^-}'}
    =\Bigl(\begin{array}{cc}
                S & T \\ -\overline{T}  &  -\overline{S}
                   \end{array}\Bigr)
 {u^+\choose u^-}
$$
 Here 
$S$, $T$, $\overline{T}$, $\overline{S}$ 
 stand for the operators associated with the matrices  
$\{s_{kl}\}_{kl}$, 
$\{t_{kl}\}_{kl}$, 
$\{\overline{t_{kl}}\}_{kl}$, 
$\{\overline{s_{kl}}\}_{kl}$ : 
$$
(Su)_k={\sum}_l s_{k\,l}u_l\;,\;(Tv)_k={\sum}_l t_{k\,l}v_l \;, \mbox{ etc. } 
$$
 Note
$$
S^*=S\;;\;T^*=\overline{T}\;;\;\overline{S}^*=\overline{S}
$$

 Next, we observe that 
$$
 [A(\cdots),A(\cdots)] 
$$
 is scalar-valued
 (up to the multiplier $I = $ the unity of the *-algebra):
$$
 [A({u^+}'\oplus {u^-}')^*,A(u^+\oplus u^-)] 
                  = <{u^+}',u^+>_0-<{u^-}',u^->_0 
$$
  where 
$<u',u>_0$
 stands for the usual inner product in
${\bf C}^N$ :
$$
 <u',u>_0
 := \overline{u'_1}u_1 + \overline{u'_2}u_2+\cdots +\overline{u'_N}u_N  
$$ 

 Motivated by this, we define:
$$
\begin{array}{rcl}
 <{u^+}'\oplus {u^-}', u^+\oplus u^-> 
 & := & 
 [A({u^+}'\oplus {u^-}')^*,A(u^+\oplus u^-)] \\
\end{array}
$$
 and we note that 
$ <{u^+}'\oplus {u^-}', u^+\oplus u^-> $ 
 is an indefinite inner product 
 on the ``space of coefficients'' of $a^*, a$.

 One can write:

$$
 <{u^+}'\oplus {u^-}', u^+\oplus u^-> 
  = ({u^+}'\oplus {u^-}', J_{a^*a}u^+\oplus u^-)
$$ 
 where one sets:
$$
 J_{a^*a} =\Bigl(\begin{array}{cc} I&0\\0&-I \end{array}\Bigr)
$$

\bigskip 

 Similar formulae hold for the 
{\bf Heisenberg-Dirac form of CCRs}: 
$$
 i[\PP_k,\QQ_l]=\delta_{k\,l}\,;\,
 i[\PP_k,\PP_l]=0\,;\,     i[\QQ_k,\QQ_l]=0\,; \, 
 {\QQ }^* =\QQ \,; \, {\PP }^* =\PP \,. 
$$
 We will write
$$
 F(x_{p}\oplus x_{q}):=x_{p1} \PP_1 + \cdots +x_{pN} \PP_N
             +x_{q1} \QQ_1 + \cdots +x_{qN} \QQ_N \,.
$$
 We use this notation whether 
$x_p\oplus x_q$ 
 is a real-valued vector or complex-valued.
 In addition, we put 
$$
 \PP (x_{p}):=x_{p1} \PP_1 + \cdots +x_{pN} \PP_N
 \quad 
 \QQ (x_{q}):=x_{q1} \QQ_1 + \cdots +x_{qN} \QQ_N \,. 
$$

 The quadratic Hamiltonian is now a formal expression
$$
 h:=\frac{1}{2}{\sum}_{l\,m}\left(
   M_{l\,m}\PP_l\PP_m - L_{l\,m}(\PP_l\QQ_m + \QQ_m\PP_l) +K_{l\,m}\QQ_l\QQ_m 
                         \right)
$$ 
 with
$M^T=M\,,K^T=K$.
 Then  
$$
 i[h,F(x_p\oplus x_q)]=F(x_{p}'\oplus x_{q}')
$$
 where 
$$
 {x_{p}' \choose x_{q}'} 
     = \Bigl(\begin{array}{cc}
             L&M\\-K&-L^T \end{array}  \Bigr) {x_{p}\choose x_{q}} 
$$

 Next, we observe that formally  
$$
 -i[F(x_{p}'\oplus x_{q}'),F(x_{p}\oplus x_{q})] 
                               = -(x_{p}',x_{q})_0+(x_{q}',x_{p})_0 
$$
 where 
$(\cdot,\cdot)_0$
 stands for the usual Euclidian-like inner product, alias 
 scalar product of vectors:
$$
 (x',x)_0:= 
 {x'_1}x_1 + {x'_2}x_2+\cdots +{x'_N}x_N  .
$$ 

 So, if we define 
$$
\begin{array}{rcl}
 s(x_{p}'\oplus x_{q}', x_{p}\oplus x_{q}) 
 &:=&
 -i[F(x_{p}'\oplus x_{q}'),F(x_{p}\oplus x_{q})] 
\end{array} 
$$
 then the
$s$ 
 becomes a
$\C $-symplectic form,
 i.e.,
 a bilinear anti-symmetric form on a complex space.

 One can write also:

$$
 s(x_{p}'\oplus x_{q}', x_{p}\oplus x_{q}) 
 = (x_{p}'\oplus x_{q}', J_{pq}x_{p}\oplus x_{q})
$$ 
 where one sets:
$$
 J_{pq} :=\Bigl(\begin{array}{cc} 0&-I\\I&0 \end{array}\Bigr)
$$

 The link between
$\am \,,\,\ap$ 
 and $\PP \,,\,\QQ $ is this:
$$
 \am=\frac{1}{\sqrt 2}(\QQ +i\PP )\,,\,\ap=\frac{1}{\sqrt 2}(\QQ -i\PP )
 \,,\,
 \QQ =\frac{1}{\sqrt 2}(\ap + \am )\,,\, \PP =\frac{i}{\sqrt 2}(\ap - \am )
$$

 Hence, 
$$
 A(u^+\oplus u^-)= \frac{1}{\sqrt 2}F(-iu^+ +iu^- \oplus u^+ +u^-)
$$
$$
 F(x_{p}\oplus x_{q})= \frac{1}{\sqrt 2}A(ix_p+x_q \oplus -ix_p+x_q)
$$

 The formal calculations show
\footnote { see Appendix A }
 that
$$
\begin{array}{rcl}
 e^{iF(x_{p}\oplus x_{q})}
 e^{iF(x_{p}'\oplus x_{q}')}
 &=& 
 e^{[iF(x_{p}\oplus x_{q}), iF(x_{p}'\oplus x_{q}')]/2}
 e^{iF(x_{p}+x_{p}'\oplus x_{q}+x_{q}')}
 \\&=& 
 e^{-[F(x_{p}\oplus x_{q}), F(x_{p}'\oplus x_{q}')]/2}
 e^{iF(x_{p}+x_{p}'\oplus x_{q}+x_{q}')}
 \\&=& 
 e^{-is(x_{p}\oplus x_{q}, x_{p}'\oplus x_{q}')/2}
 e^{iF(x_{p}+x_{p}'\oplus x_{q}+x_{q}')} \,,
 \\
\Big( e^{iF(x_{p}\oplus x_{q})} \Big)^*
 &=& 
 e^{ -iF(\overline{x_{p}}\oplus \overline{x_{q}})} \,.
\end{array}
$$
 These formulae are the so-called {\bf exponential form of the CCRs}.
 We adopt these formulae as a tenet, as an axiom.

 If one restricts himself to the case of real-valued
$x_p, x_q$, then one has the relations  

$$
\begin{array}{rcl}
 e^{iF(x_{p}\oplus x_{q})}
 e^{iF(x_{p}'\oplus x_{q}')}
 &=& 
 e^{-is(x_{p}\oplus x_{q}, x_{p}'\oplus x_{q}')/2}
 e^{iF(x_{p}+x_{p}'\oplus x_{q}+x_{q}')} \,,
 \\
\Big( e^{iF(x_{p}\oplus x_{q})} \Big)^*
 &=& 
 e^{ -iF(x_{p}\oplus x_{q})} \,.
\end{array}
$$

 This form of the CCRs is called {\bf Weyl}.
\footnote{In this case, $e^{iF(x_{p}\oplus x_{q})}$ is unitary,}
\footnote{ 
 and $s(\cdot,\cdot)$
 is the usual (pre)symplectic form, i.e., the real-valued
 anti-symmetric form on real space}
 
 If a Hamiltonian has been given,
 the standard quantum mechanics practice suggests
 solving  
 the dynamical equations, in particular
$$
 \frac{\partial F(x_p\oplus x_q)(t)}{\partial t}
       =i[h,F(x_p\oplus x_q)(t)] 
$$
 or ``equivalently'' 
$$
 \frac{\partial A({u^+}\oplus {u^-})(t)}{\partial t} 
      = i[h,A(u^+\oplus u^-)(t)] 
$$
 A formal calculation allows one to rewrite this equation as 
$$
 \frac{\partial }{\partial t }
   {x_{p}(t) \choose x_{q}(t)} 
       = \Bigl(\begin{array}{cc}
              L&M\\-K&-L^T \end{array}  \Bigr)
   {x_{p}(t)\choose x_{q}(t)} 
$$
 respectively
$$ 
 \frac{\partial }{\partial t} 
   {{u^+}(t)\choose {u^-}(t)} 
     =  i\Bigr(\begin{array}{cc}
            S & T \\ -\overline{T}  &  -\overline{S}
              \end{array}\Bigr) 
 {u^+(t)\choose u^-(t)}
$$
 We will suppose, the solution to the equation exists and 
 let 
$V(t,s)$ 
 denote the corresponding propagator.
\footnote{ it means evolution operator}

 Then,   
$V(t,s)$ 
 lifts to an *-automorphism  
$\alpha_{V(t,s)}$ 
 of CCRs by
$$
 \alpha_{V(t,s)}F(x_p\oplus x_q)
     := F(V(t,s)(x_p\oplus x_q))
$$
$$ 
 \alpha_{V(t,s)} e^{iF(x_{p}\oplus x_{q})}
      := e^{iF(V(t,s)(x_{p}\oplus x_{q}))}
$$ 
 This *-automorphism is called either
 the {\bf linear canonical transformation}
 or
 {\bf Bogoliubov transformation}
 or
 {\bf quasi-free }
 automorphism of a Bose system.

 Given a ``physical system'' state 
$\omega$, e.g., whether a ground state or  
 a state with an interesting energy distribution 
 or a state with the momentum at given exact value (``plane wave state'')
 or something like that, 
 and given an observable 
${\bf A}$, i.e., something like a momentum, energy, position, spin, 
 particles number, etc,,
 we will denote the corresponding 
 {\bf expectation value of 
${\bf A}$ 
 at the state 
$\omega$  } 
 by 
$\omega{\bf A}$. 
 The functionals 
$$
 x_{p}\oplus x_{q} \mapsto \omega e^{ iF(x_{p}\oplus x_{q})}
$$
 and 
$$
  u^+\oplus u^-  \mapsto \omega e^{ A(u^+\oplus u^-) } 
$$
 are called the 
 {\bf characteristic functionals} or {\bf characteristic functions} 
 of the state 
$\omega $. 
 If the function 
$$
 \lambda\in\R   
 \mapsto \omega 
 e^{ iF(x_{p}^0\oplus x_{q}^0+\lambda x_{p}\oplus x_{q})}
$$
 is continuous, 
 then the state 
$\omega $  
 is called {\bf regular }. As a rule, one assumes therewith that 
 the coefficients 
$x_{p}^0\oplus x_{q}^0, x_{p}\oplus x_{q}$  
 are real-valued. 
 If 
$$
\omega e^{ iF(x_{p}\oplus x_{q})} 
 = e^{- \mbox{ quadratic function of } x_{p}\oplus x_{q}}
$$
 then 
$\omega$ 
 is called {\bf even quasi-free}.

 We will slightly extend the class of these states
 and choose a definition of the {\bf even quasi-free-like } states 
 which emphasizes the latter, algebraic, property 
 and partially deemphasizes the continuity property.
 The main idea behind the states we will discuss is briefly this:
\footnote{ a precise definition see in the next Section }
 Let us suppose that we consider a one-dimensional system  
 and even quasi-free states given by 
$$
\omega e^{ iF(x_{p}\oplus x_{q})} 
 =\omega e^{\displaystyle ix_{p}\PP + ix_{q}\QQ } 
 = e^{\displaystyle - \frac{1}{4}(a x_{p}^2 + b x_{q}^2)}\,,\quad, a>0, b>0
$$

 The uncertainty principle prescribes 
$ab \geq 1$, 
 and therefore we could not assign 
$a:=0$,
 whatever real number  $b$ we had chosen.
 But why not 
$a:=0, b:=+\infty$?'' 
 i.e., why not 
$$
 \omega e^{\displaystyle ix_{p}\PP + ix_{q}\QQ }
 :=\left\{\begin{array}{cl}
         1,& \mbox{ if } x_{q} =0 \\
         0,& \mbox{ if } x_{q} \not=0 \\
         \end{array}\right. 
$$
 Or why not?'':
$$
 \omega e^{\displaystyle ix_{p}\PP + ix_{q}\QQ }
 :=\left\{\begin{array}{cl}
         e^{-b x_{q}^2/4},& \mbox{ if } x_{p} =0 \\
         0,             & \mbox{ if } x_{p} \not=0 \\
         \end{array}\right. 
$$
 Actually, we may take 
$\epsilon >0 $ and define states
$\omega_{\epsilon}$
 by 
$$
 \omega_{\epsilon}e^{\displaystyle ix_{p}\PP + ix_{q}\QQ }
 := e^{\displaystyle - \frac{1}{4}
 \left( x_{p}^2 /{\epsilon}+ (b+\epsilon)x_{q}^2\right)} 
 \,,\quad  b\geq 0
$$
 Then we may take limit of 
$\omega_{\epsilon}$ 
 as 
$\epsilon \to +0$ 
 without loss of the main 
 algebraic property of states being positive definite, 
 although we lose partially the continuity property. 
 Thus we obtain just 
$$
 \omega_{+0}e^{\displaystyle ix_{p}\PP + ix_{q}\QQ }
 := \lim_{\epsilon\to +0}
 \omega_{\epsilon}e^{\displaystyle ix_{p}\PP + ix_{q}\QQ }
 =\left\{\begin{array}{cl}
         e^{-b x_{q}^2/4},& \mbox{ if } x_{p} =0 \\
         0,               & \mbox{ if } x_{p} \not=0 \\
         \end{array}\right. 
$$
 and we can refer to such states as abstractions of the usual states, 
 perhaps as artificial abstractions. 
 These states as well as any $N$-dimensional analogue,
 we will call such states the 
 {\bf even quasi-free-like} or {\bf quadratic} states. 
 
 It's our object. 

\begin{Thm}{COMMENT}{ } \rm 

    The more detailed description of the notions in this section,  
 one can find in e.g.,  [Ber], [BR2],
 and especially [Fey, Stat.Mech.].
  Notice that our definition of 
$\PP$ and $\QQ$ slightly  differs from the standard. 
 As a rule one sets 
$\PP$ and $\QQ$ 
 so as to  
$$
      a=\frac{1}{\sqrt 2}(\QQ +i\PP )\,,\,a^*=\frac{1}{\sqrt 2}(\QQ -i\PP )
      \,,\,
      \QQ =\frac{1}{\sqrt 2}(a^* + a )\,,\, \PP =\frac{i}{\sqrt 2}(a^* - a )
$$

\end{Thm} 

\newpage
\section{Quadratic and Quasi-free States on CCRs Algebra and Quasi-Free-like
 Atomorphisms }

  In this section we discuss notions 
 related to that introduced in the previous section.

 First, we give an abstract axiomatic definition of the CCRs.

 Let 
$Z$ 
 be a real vector space, 
$s:Z\times Z\to R$ 
 a bilinear antisymmetric form
($s$ need not be nondegenerated)
 and 
$V:Z\to Z$ 
 be a linear operator. 

  Let 
$\KK , <,>, V_{\C }$
 be the standard complexification of 
$Z, is(,), V $, 
 i.e. 
$\KK $ 
 is the standard 
$\R $-linear doublication of 
$Z$ 
$$
 \KK := {\C }Z = Z\oplus_{\R }Z  \,,
$$
 the multiplication with $i$ is given by
$$
 i(\f \oplus \g ) := (-\g )\oplus \f  \,,
$$
 and $<,>$  is the standard sesquilinear 
 extention of 
$is(,)$ and $V_{\C }$ is the standard $\C $-linear extention of $V$ .

 Often, we will write 
$$\f +i\g \mbox{ instead of }  \f \oplus \g  $$
 and next the symbol $C$ will denote the natural complex conjugation in 
$\KK $ :
$$ C(\f +i\g ) := \f -i\g 
\,, \qquad (\f ,\g \in Z)$$

 So, 
$\KK ,<,> $ is indefinite inner product space with $<,>$-antiunitary
 involution $C$:
$$ C^2 =I\,, \qquad   <C\f ,C\g > = <\g ,\f >
\,, \qquad (\f ,\g \in Z)$$

\begin{Remark}{ 1.} 

 We take it as known that
$V$ is a homomorphism of $s(,)$,
 i.e.
$$  s(V\f ,V\g )=s(\f ,\g )  \qquad(\forall \f ,\g \in Z)$$
 iff  
 and $V_{\C }$ is a $<,>$-isometric operator. 
 Similarly $V$ is an automorphism of $s(,)$
 i.e.
$$ 
 V \mbox{ is bijective and } 
 s(V\f ,V\g )=s(\f ,\g ) \qquad (\forall \f ,\g \in Z)$$
 iff 
$V_{\C }$ is a $<,>$-unitary operator. 
\end{Remark}

\begin{Definition}{ 1.} 

 {\bf Abstract Weyl  *-algebra}, 
 we denote it by $W_{Z,s}$ , is here a free *-algebra
 on the symbols $\e{\f }, \f \in Z$ subject to the relations
$$
 (\e{\f })^*=\e{-\f } \;,\; 
 \e{\f }\e{\g } \;=\;
 e^{-is(\f ,\g )/2}\e{\f +\g }\;.
$$
\end{Definition}

\begin{Remark}{ 2. {\rm(cf. e.g.,  [MV],[BR])}} 

    If V is an automorphism of 
$s$,
 then the correspondence     
$\epsilon^{\f }\mapsto \epsilon^{V\f }$
 induces 
 a *-automorphism; this *-automorphism
 is called {\bf quasi-free}, often, {\bf Bogoliubov *-automorphism },
 alias {\bf Bogoliubov transformation}.
 We will denote it by $\alpha_V$ 

 If 
$\chi$ 
 is a *-character of the additive group 
$Z$
 i.e. if
$$
 \chi(\f +\g )= \chi(\f )\chi( \g ),  \chi(\f )^*=\chi(-\f ), 
 \forall \f , \g \in Z
$$
 then the correspondence     
$\epsilon^{\f }\mapsto \chi(\f )\epsilon^{V\f }$
 extends to a {\bf gauge-like *-automorphism}, 
 alias {\bf coherent *-automorphism}; 
 we denote it by $\alpha_{\chi}$ .

 If 
$$ \chi(\f ) = e^{il(\f )} $$
 where 
$l$ 
 is a real-valued 
$mod2\pi$-additive function on
$Z$,
 then we prefer to write 
$\alpha_l$
 instead of
$\alpha_{\chi}$.

 The automorphisms of the form 
$$
 \alpha_{V,\chi} := \alpha_{V}\alpha_{\chi},
\qquad
 (\alpha_{\chi,V} := \alpha_{\chi}\alpha_{V})
$$
 are called {\bf quasi-free-like}.
\end{Remark}


 The rest of this section until Example 1 
 is an extending modificaton
 of the Manuceau Verbeure Theory
 of quasi-free states.

 It will be convenient to change (equivalently!)  
 the usual definition of positive quadratic form. 

\begin{Definition}{ 2.}

  We will say that
$q:Z\to [0,\infty]$ 
  is  {\bf quadratic} iff
$$
    q(\f +\g )+q(\f -\g )=2[q(\f )+q(\g )]
$$
$$
    q(k\f )\,=\,k^2q(\f ) \quad (\f ,\g \in Z,\;k\in R)
$$
$$
\mbox{(hereafter }
   0\cdot \infty \,=\,0\;,\;\infty \,+\,\infty \,=\,\infty \;
\mbox{and so on})
$$
  
 The set 
$$
 Q(q) := \{\f \in Z| q(\f ) < \infty\} 
      \equiv \{\f \in Z| q(\f ) \not= \infty\} 
$$
 is called the {\bf form domain} or the {\bf domain} of  
$q$. 
 If 
$Q(q) = Z$, then 
$q$ is called {\bf finite}. 
 
 Given two quadratic $q_1, \ q_2$, we write
$$
  q_1 \leq q_2  \mbox{\qquad  iff \qquad}  
  q_1(\f ) \leq q_2(\f )    \qquad (\forall \f \in Z)
$$
\end{Definition}

\begin{Remark}{ 3.} 

 Given a quadratic
$q$, 
 the form domain of 
$q$ 
 is linear, and
$q$ 
 is associated with a unique symmetric bilinear positive 
 (if $Z$ is over $\R $)
 or symmetric sesquilinear positive form
 (if $Z$ is over $\C $),
 we denote it by
$q(\cdot,\cdot)$;
 this form can be recovered from the
$q$ 
 by the {\bf polarization identity}
$$
 q(\f , \g )
 = \frac{1}{2}(q(\f +\g ) + q(\f -\g ))  
 \qquad\mbox{\rm (if $Z$ is over $\R $) }
$$ 
 resp.
$$
 q(\f , \g ) = \frac{1}{4}
 (q(\f +\g ) + q(\f -\g ) -iq(\f +i\g ) + iq(\f -i\g )) 
 \qquad\mbox{\rm (if $Z$ is over $\C $) }
$$ 

\end{Remark}

\begin{Definition}{ 3.}

  We say  $q$ is a quadratic-like {\bf majorant of $s$},  iff
$$
         2|s(\f ,\g )| \le q(\f ) + q(\g ) \qquad (\f ,\g \in Z)
$$
 and if, of course, 
$q$ 
 in itself is quadratic.  

\end{Definition}
\begin{Proposition}{ 1.} 

 For any majorant
$q$,  
 there exists a minimal quadratic-like majorant, say 
$q_0$,
 such that  
$q_0 \leq q$. 
 Hereafter, we mean by 
`$q_0 $
 is a 
 {\bf minimal majorant}' that, 
 if 
$q_1 \leq q_0 $ 
 for a quadratic majorant   
$q_1$, 
 then  
$q_1 =q_0 $. 
\end{Proposition}

\begin{Definition}{ 4.{\rm cf. [Oks])}} 

 We say that a linear *-functional $\omega $ on $W_{Z,s}$ is {\bf quadratic
 (alias even quasi-free-like, generalized even quasi-free)} iff
$$
  \omega \epsilon^{\f } = e^{-q(\f )/4} \quad (e^{-\infty }\,=\,0)
$$
 for a quadratic $q$ .
\end{Definition}

\begin{Theorem}{ 1.}

 {\rm (i)}
\hfill \parbox[t]{.9\textwidth}{
  A quadratic 
$\omega$ is a state iff the associated 
$q$ 
 is a quadratic-like majorant of $s$.
}

 {\rm (ii)}
\hfill \parbox[t]{.9\textwidth}{
  A quadratic
$\omega$ is a pure state iff the associated
$q$
 is a minimal quadratic-like majorant of $s$.
}

\end{Theorem} 

\begin{Example}{ 1.}

 Put 
$q(\f ):=0$ 
 at 
$\f =0$ 
 and
$q(\f ):=+\infty $
 otherwise, 
 i.e., define a linear functional 
$\delta_0$
 on 
$W_{Z,s}$
 so that 
$$
 \delta_0\epsilon^{\f } 
    := \cases{1,  &  if  $\f = 0$ \cr  0, & if $\f \not=0$ } 
$$

 Then, the  
$q$
 is a quadratic-like majorant, called {\bf trivial },
 and 
$\delta_0$ 
 is {\bf trivial } state. 
 Notice {\rm (e.g.,  [BR2, p.79], EXAMPLE 5.3.2)}, the   
$\delta_0$
 is a trace-state on
$W_{Z,s}$. 
\footnote{ and this is the unique trace-state if 
$s(\cdot ,\cdot )$ is nondegenerate}
 In addition, this state is invariant under all Bogoliubov transformations.

\end{Example}

\begin{Definition}{ 5.} 

 Given a quadratic $q$, we denote its standard complexification
 by 
$q_{\C}$. 
 We define it so:
$$
 q_{\C}(\f +i\g ) := q(\f )+q(\g )
$$
 A similar notation is given to the complexification 
 of an arbitrary linear ${\cal T} : \, Z\to Z$:
$$
 {\cal T}_{\C}(\f +i\g ) := {\cal T}\f + i{\cal T}\g  \,.
$$
\end{Definition}

\begin{Remark}{ 4.} 

 For the complex space case the definition of the sentence
 `a functional, e.g., $q_{\C}$, is quadratic'
 is to be modified:
 $$ q_{\C}(kz)=|k|^2q_{\C}(z) \,.$$
 The rest of the definition remains as befor.
\end{Remark}

\begin{Observation}{ 1.}

 Given a linear $\tilde {\cal T}: \KK \to \KK $, it is of the form 
$\tilde {\cal T}= {\cal T}_{\C}$
 for a suitable ${\cal T} : Z\to Z$ iff
$$ C\tilde {\cal T}C = \tilde {\cal T}. $$
 Similarly, for any quadratic $\tilde q : \KK \to [0,\infty]$, 
 there is a quadratic $q: Z \to [0,\infty]$
 such that $\tilde q =q_{\C}$
 iff 
$$ \tilde qC= \tilde q $$
\end{Observation}

\begin{Theorem}{ 2.}

 {\rm (i)}
\hfill \parbox[t]{.9\textwidth}{
$q$
 is a quadratic-like majorant of $s$
 iff
$q_{\C}$
 is a quadratic-like majorant of $<,>$
}

 {\rm (ii)}
\hfill \parbox[t]{.9\textwidth}{
$q$
 is a minimal quadratic-like majorant of $s$
 iff
$q_{\C}$
 is a minimal quadratic-like majorant of $<,>$
}

\end{Theorem} 

\begin{Remark}{ 5.} 

  We will deal, first and foremost, with {\em quadratic-like} majorants.  
  So, if no confusion can ocurr, we will omit the particle `-like'
  or the whole word `quadratic-like', 
 although occasionally we will repeat the whole term 
 `quadratic-like majorant' 
 for emphasis. 

 \end{Remark}  

 If a new object is declared,
 the first question is whether this object 
 does exist.
 Of course, automorphisms, majorants and invariant 
 (under Bogoliubov transformation) quadratic states,
 all these objects do exist.   
 Interestingly enough, 
 a {\bf finite } majorant {\bf need not} exist;
 a quasi-free *-automorphism 
 {\bf need not} 
 have an { \bf invariant non-trivial} quadratic state.

\begin{Example}{ 2. {\rm ([Bog, p.62-63, {\bf Example 3.2}])}}

 Let 
$\HH $ be the vector space of those doubly infinite numerical sequences 
 where only a finite number of terms with negative index is different 
 from zero, and for 
$\f =\{\xi_j\}_{j\in {\bf Z}} \in \HH $, 
$\g =\{\eta_j\}_{j\in {\bf Z}} \in \HH $
 let
$$
 <\f ,\g > := \sum_{j=-\infty}^{\infty}\overline{\xi_{j}} \eta_{-j-1}
$$
 Then 
$<\cdot,\cdot>$
 cannot have a norm majorant.
\footnote{ consequently, 
 cannot have a finite quadratic majorant }  
\end{Example}

 We are interested in the symplectic space case 
 and translate the previous Example 2 into obvious:

\begin{Example}{ 3.}

 For this Example, let
${\bf S_0}$ 
 denote the linear space of those real-valued sequences 
$ \f : {\bf N} \to {\bf R}$
 such that 
$$
 \f (n) =0  
 \mbox{ \rm for all but a finite number of } 
 n 
$$
 and 
${\bf S_{all}}$
 denote the linear space of all real-valued sequences.
 Finally, define 
$$
 Z := {\bf S_0}\oplus {\bf S_{all}} 
$$
 and 
$$
 s(\f_1 \oplus \f_2, \g_1 \oplus \g_2)
 := \sum_{n}
  (\f_1(n)  \g_2(n) -  \f_2(n)  \g_1(n))
$$
 Then the symplectic form  
$s(\cdot,\cdot)$
 cannot have a norm 
 majorant. 
\end{Example}

\begin{Example}{ 4. {\rm ([Ch4])}} 

 Let 
$\bf R[Z]$ 
 be the free real *-algebra on the symbols 
$u[n], n \in {\bf Z}$ 
 subject to the relations 
\footnote{ we will deal with the so-called group *-algebra of
 {\bf Z} over {\bf R} } 
$$
 u[n]u[m] := u[n+m], u[n]^* := u[-n]   \qquad (n,m \in {\bf R[Z]} ) 
\quad
\footnote{ thus $u[n]^*u[n]=u[n]u[n]^* = u[0] =1$  } 
$$
 Let 
$f$ 
 be a linear functional defined by 
$$ 
 fu[n] := e^{\sqrt{|n|}} - e^{-\sqrt{|n|}} 
$$ 
 Next, define 
$$ 
 K_f 
 := \{K\in {\bf R[Z]}| \quad (\forall A\in {\bf R[Z]}) \quad f(A^*K) = 0 \} 
$$ 
 and 
$$ 
 A_f := A + K_f  \qquad (A \in {\bf R[Z]}) 
\quad 
\footnote{ actually, 
$K_f =\{0\}$ 
 and in essence 
$A_f= A$, 
 however we ignore it } 
$$

 Then, the bilinear anti-symmetric form 
$$
 A,B \mapsto f(A^*B)
$$ 
 lifts to a symplectic form, 
$s$,
 on the quotient space 
$ {\bf R[Z]}/K_f $ 
 and the map 
$$
 A\in {\bf R[Z]} \mapsto u[1]A \in {\bf R[Z]}  \qquad  (A\in {\bf R[Z]})
$$
 lifts to a symplectic automorphism 
$V :  {\bf R[Z]}/K_f \to  {\bf R[Z]}/K_f $ ;
 they are correctly defined by
$$
 s(A_f,B_f):= f(A^*B)  \qquad (A, B \in {\bf R[Z]} ) 
$$ 
$$
 V A_f := (u[1]A)_f   \qquad ( A\in {\bf R[Z]} )
$$ 
 
 Now then, there is no non-trivial 
$V$-invariant majorant of $s$
 and there is no non-trivial quadratic 
$\alpha_V$-invariant 
 state on 
$W_{Z,s}$ 
 where  
$ Z:={\bf R[Z]}/K_f $.  
 
\end{Example} 

\begin{Thm}{COMMENT}{ } \rm 

 A state 
$\omega$ 
 is said to be regular iff 
 the function 
$ x \in {\R} \to \omega \epsilon^{x\f +\g}$ 
 is continuous 
 whatever 
$\f $ and $\g $. 
 
 Manuceau and Verbeure [MV] discussed only regular 
 states and therefore only {\em finite } quadratic forms
 and the corresponding states.
 As for non-regular states, one can confer the approach in this section
 with one of [FS], [Gru], [LMS], [CMS], 
 and especially with that of [Oks]. 
 Recently Halvorson [Hal] proposed a very interesting standpoint
 which is reminiscent of some of the papers 
 of Antonets, Shereshevski, first of all [AS].

 The `non-regular' part of this section 
 is based entirely on [Ch1], [Ch2], [Ch3] and the summarizing [Ch4]. 

 In Example 4, we have applied a GNS-like 
\footnote{ GNS = Gelfand--Naimark--Segal }  
 construction. 
  For detailes of such constructions, 
  see e.g., [Schatz] or/and [BD] 
 (or [Ch 2--4 ], if one deals
 with 
 the objects discussed in this Section). 

\end{Thm}

\newpage

\section{The Case of Regular Spaces }

 In the previous section we discussed
 relatively general spaces and forms.
 So, the statements were `in general'. 

 With stronger hypothesis on 
$Z$, $\KK $ 
 and forms 
$s(\cdot,\cdot)$, $<\cdot,\cdot>$ 
 one can obtain a stronger conclusion.
%
 We start with two restricting
 definitions
 which one finds among the primary definitions of two different theories.
 We mean the standard theory of the quasi-free states (e.g., [BR2]) 
 and, as for the second definition,
 the so-called Krein spaces theory (e.g., [Bog])

\begin{Definition}{ 1.} 

 $Z,\,  s $ is said to be {\bf regular} iff there is a linear
 $J: Z\to Z$ such that

 1)  $s(J\f , J\g ) =s(\f ,\g )  \qquad    \forall  \f ,\g \in Z$;

 2)  $J^2=-I$;

 3)  $s(\f ,J\f ) \geq 0    \qquad  \forall \f \in Z$ ,

 4)  $Z$ is a real pre-Hilbert space with respect to the  
     scalar product  
     $\f ,\g \to  s(\f ,J\g )  \qquad (\f ,\g \in Z)$ .
\end{Definition}

\begin{Definition}{ 2. {(e.g., [Bog])}} 

 Let 
${\cal K} ,\,, <,>  $
 be an inner product space.

${\cal K}, \,, <,>  $
 is said to be 
 {\bf regular indefinite inner product space}
 iff there is a linear
$\JJ : {\cal K} \to {\cal K} $ such that

1)  $<\JJ z, \JJ w> =<z,w>    \qquad   \forall  z,w\in {\cal K} $;

2)  $\JJ ^2=I$;

3)  $<z,\JJ z> \geq 0    \qquad  \forall z\in {\cal K} $ ,

4)  ${\cal K}$ is a pre-Hilbert space with respect to the  
    scalar product  
    \mbox{ $z,w \to  <z,Jw>  \qquad (z,w \in {\cal K} )$ . }

 If 
${\cal K}$ 
 is complete,
 then
${\cal K}$ 
 is said to be 
 {\bf a Krein space}. 
\end{Definition}

 We see one definition is very much like another. 
 We state
 it in the mathematical terms as an 

\begin{Observation}{ 1.}

 $Z,\,s $  is regular if and only if the corresponding
 standard complexification of 
$Z,\,s $,
 i.e.,
$\KK ,\,<,>$ in the sense of the previous section, 
 is a regular indefinite inner product space.
 For the corresponding $\JJ$, we may take 
$$\JJ:=i J_{\C}\,,\,
\mbox{\rm recall that } 
 J_{\C}:= \mbox{ standard complexification of } J \,.$$
\end{Observation}

 The idea behind the constructions we will discuss is very simple. 
 If we see that some of the primary definitions 
 of two different theories are similar,
 then we expect it may be well worth stating   
 the similarity between the results of these theories.
 Thus we need to elaborate 
 a machinery 
 so that we could translate the statements of the one theory 
 into the language of another. 

 First, we consider what is common in both 
 languages 
 and we start to do it 
 by introducing the general notations of the basic terms. 
          

 The symbol \HH\ will denote a linear space, real or complex,
 and $b$ be a bilinear or sesquilinear form respectively.
 In addition, we suppose that $b$ is whether symmetric (hermitian for $\C$)
 or antisymmetric (antihermitian for $\C$).
  
 The symbol 
$\J$ 
 will denote a linear operator which has either the properties
$$ \begin{array}{cc} \J^2 :=-I\\     
   b(\J \f ,\J \g ) = b(\f ,\g )\end{array}  
  \qquad\mbox{ case of antisymmetric (antihermitan) $b$ ,}     $$ 
 or
$$ \begin{array}{cc} \J^2 :=I\\     
   b(\J z,\J w) = b(z,w)\end{array}  
  \qquad\mbox{ case of symmetric (hermitian) $b$.}    $$ 
 In addition, we define, unless otherwise specified, that
$$  \J^* :=-\J  \qquad\mbox{ case of antisymmetric (antihermitian) $b$ ,}  $$ 
$$  \J^* :=\J  \qquad\mbox{ case of symmetric (hermitan) $b$ .}    $$ 
 It is unlikely 
 that this definition can produce any confusion:
 we will deal, typically,  
 with {\em nondegenerated } forms $s \,,\, <,> $;
 in these cases $\J^*$ will coincide with standard
$s$-  or  $<,>-$ adjoint of $\J$ respectively.


 There are two classes 
 of regular spaces which we will discuss.
 The first class is given by:

\begin{Example}{ 1.}

 Let
$Z_0$ 
 be a real or complex Hilbert or pre-Hilbert
 space.
 Put
$$
 Z:= Z_0\oplus Z_0
$$
 and
$$
 \J :=  \left( \begin{array}{cc}0&-I\\I&0 \end{array}  \right)
$$
 This choice of
$Z$
 and 
$ \J $
 corresponds to
%
 the case where we adopt a definition of CCRs phrased in terms of 
$\PP ,\QQ $, i.e., in terms of momentum and position operators.
%
%
\end{Example}

 Another class 
 of regular spaces is:

\begin{Example}{ 2.}

 Let
$H_0$
 be a complex Hilbert or pre-Hilbert space.
 Put   
$H_+ := H_0$, $H_- := H_0$ , 
$$
 \KK := H_+ \oplus H_-
$$
 and 
$$
 \J :=  \left( \begin{array}{cc}I&0\\0&-I \end{array}  \right)
$$
 This case corresponds to
 that, when one adapts himself to the CCRs phrased in terms of 
$a^* ,a $, i.e., in terms of creation and annihilation operators.
\end{Example}

 The connection between these classes is simple:
%

\begin{Observation}{ 2.}

 If 
$$ 
 J_{a^*a}
 := 
\left(
\begin{array}{cc}
  1 &  0 \\ 
  0 & -1
\end{array}
\right)
 \,,\quad
 iJ_{pq}
 := 
\left(
\begin{array}{cc}
  0 & -i \\ 
  i &  0
\end{array}
\right)
$$

 then 
$$ 
 J_{a^*a}
\left(
\begin{array}{cc}
  i & 1 \\ 
 -i & 1
\end{array}
\right)
 =
\left(
\begin{array}{cc}
 i &  1 \\ 
 i & -1
\end{array}
\right)
 =
\left(
\begin{array}{cc}
  i & 1 \\ 
 -i & 1
\end{array}
\right)
 iJ_{pq} 
$$ 
$$
 iJ_{pq}
\left(
\begin{array}{cc}
 -i & i \\ 
  1 & 1
\end{array}
\right)
 =
\left(
\begin{array}{cc}
 -i &  i \\ 
  1 & -1
\end{array}
\right)
 =
\left(
\begin{array}{cc}
 -i & i \\ 
  1 & 1
\end{array}
\right)
 J_{a^*a}
$$
$$
\left(
\begin{array}{cc}
  i & 1 \\ 
 -i & 1
\end{array}
\right)
\left(
\begin{array}{cc}
 -i & i \\ 
  1 & 1
\end{array}
\right)
 = 
\left(
\begin{array}{cc}
  2  &  0 \\ 
  0  &  2
\end{array}
\right)
 = 
\left(
\begin{array}{cc}
 -i & i \\ 
  1 & 1
\end{array}
\right)
\left(
\begin{array}{cc}
  i & 1 \\ 
 -i & 1
\end{array}
\right)
$$

\end{Observation}


 We now return to 
quadratic forms and majorants 
 and
 henceforth in
 this section we {\bf assume that 
 the above spaces 
$Z_0$ 
 and 
$H_0$ 
 are complete.} 
 
\begin{Definition}{ 3.} 
 
 We say  
 a form 
$q$ 
 is {\bf closed} if 
$q$ 
 is closed as a usual quadratic form on Hilbert space 
$$
 \overline{Q(q)} 
 := \mbox{ closure of $Q(q)$ in $\HH$ with respect to $\|\cdot \|$ },
$$ 
 i.e. is closed in the sense 
 adopted in [RS1].
\end{Definition} 
\begin{Theorem}{ 1.}

 Let 
$q$ 
 be a majorant.

 If 
$q$ 
 is minimal, then  
$q$ 
 is closed.

\end{Theorem}

\begin{Definition}{ 4.}  

 We say a form  
$q$ 
 has an {\bf operator representative } if there exists an 
${\cal T}: D_{\cal T} \subset \HH \to \HH$ 
 such that 
$$
 D_T = Q(q) \mbox{ and }  
 q(\f ) = \|{\cal T}\f \|^2  \qquad (\f \in Q(q)) \,.
$$   
\end{Definition}

\begin{Theorem}{ 2.} 

 If 
$q$ 
 is closed, then  
$q$ 
 has an operator representative.
 
 In addition, there is a unique 
 self-adjoint operator 
$Q: D_Q \subset \overline{Q(q)} \to \overline{Q(q)} $ 
 such that   
$$
 D_{Q^{1/2}}=Q(q) \mbox{ and } q_Q := \|Q^{\frac12}x\|^2.
$$
 
\end{Theorem} 

\begin{Proof}{. }  
 Straightforward from Definitions 3, 4 and Theorem 1, using [RS1].
\end{Proof}

\begin{Definition}{ 5.} 

 In the situation of the Theorem 2,
 we say  
$Q$ 
 is the {\bf operator of } 
$q$
 and write
$q=q_Q$. 
 Thus, we isolate a class of majorants. 
 We will call these majorants {\bf operator majorants}.

 Let 
$Q$ be the operator of $q$.    
 Then we write

$$
P:=\mbox{ orthogonal projection of } \HH \mbox{ onto } \overline{Q(q)} ,
$$ 
$$
 R:=(I+Q)^{-1}P.
$$

\end{Definition} 

\begin{Remark}{ 1.} 

 It is evident that
$$
 0\leq R \leq I  , \qquad R=R^*
$$
 and that 
$Q$  (and  $q$)
 can be recovered from the 
$R$
 by the formulae
$$
 Q = R^{-1}-I  ; D_Q = Ran\,R \,.
$$

\end{Remark}

 We now characterize the operator majorants by means of the above
$Q$
 and
$R$. 
\begin{Theorem}{ 3.}

 {\rm (i)}
\hfill \parbox[t]{.9\textwidth}{
$q_Q$  is a majorant iff
$$
 R + \J^*R\J \leq I \,; 
$$
}

 {\rm (ii)}
\hfill \parbox[t]{.9\textwidth}{
$q_Q$  is a minimal majorant iff
$$
 R + \J^*R\J = I. 
$$
}

\end{Theorem}

 If we handle indefinite inner product space, we can say more:

\begin{Observation}{ 3.}

 Let
$\HH =\KK $ i.e., let us assume 
$\J^*=\J $.
 Then, 
$$
 R + \J^*R\J = I, \qquad 0\leq R \leq I
$$
 if and only if
$$
 R =\frac12 \left( \begin{array}{cc}I&K^*\\K&I \end{array}  \right) \,,
$$
 where 
$ 
 R \mbox{ is thought of as an operator from }
 H_+ \oplus H_- 
   \mbox{ into }   
 H_+ \oplus H_- \,,
$ 
\\ and where $K$ is an operator such that $\|K\| \leq 1$,
 or, \\ equivalently,
$$ 
 R 
  =\frac{1}{4}
\left(
\begin{array}{cc}
  2 - K - K^*  &   iK - iK^* \\ 
  iK - iK^*    &   2 + K + K^* 
\end{array}
\right)
$$
 where 
$ 
 R \mbox{ is thought of as an operator from }
 Z_0 \oplus Z_0 
   \mbox{ into }   
 Z_0 \oplus Z_0 \,,
$ 
\\ and where 
$K$ 
is the same operator as above.
\end{Observation}

 With this Observation 3, Theorem 3 implies  
\begin{Corollary}{ 1.}

 Let
$\HH =\KK $.
 Then $q_Q$  is a minimal majorant if and only if
$$
 R =\frac12 \left( \begin{array}{cc}I&K^*\\K&I \end{array}  \right)
 \mbox{ with respect to } R: H_+ \oplus H_- \to H_+ \oplus H_- 
$$
 and with a $K$ such that $ \|K\| \leq 1$ 
 or, equivalently,
$$ 
 R 
  =\frac{1}{4}
\left(
\begin{array}{cc}
  2 - K - K^*  &   iK - iK^* \\ 
  iK - iK^*    &   2 + K + K^* 
\end{array}
\right)
 \mbox{ with respect to } R: Z_0 \oplus Z_0 \to  Z_0 \oplus Z_0  
$$
 and with the same 
$K$. 
\end{Corollary}

\vfill

 Finally,
 we turn to the question:
 What about {\bf complexificated } majorants and operators?
 
 The theorems are:
\begin{Theorem}{ 4. {\rm (cf. Observation 3.1)}}

 In the situation described in Example 1, let
$$
 {\KK}_0 := {\C }Z_0 = \mbox{ \rm the standard complexification of } Z_0 
$$ 
$$
 C_0 := \mbox{ the corresponding complex conjugation operator on } {\KK}_0
 \,,\quad 
 C := C_0 \oplus C_0 
$$ 

 Then, 
 a minimal majorant 
$\tilde q $
 on 
$\KK = {\KK}_0 \oplus {\KK}_0 $ 
 is a complexification of a (minimal majorant) 
$q$ on $Z=Z_0 \oplus Z_0 $ 
 i.e., 
$\tilde q $
 is of the form 
$\tilde q = q_{\C }$ 
 if and only if 

$$ \tilde qC= \tilde q $$
 or, equivalently, 
$$R=CRC$$
 or, equivalently,
$$ K^*= \overline{K} := C_0KC_0 $$

\end{Theorem}
\begin{Theorem}{ 5. {\rm (cf. Observation 3.1)}} 

 A linear 
$\tilde {\cal T}: {\KK}_0 \oplus \to {\KK}_0 \oplus {\KK}_0  $ 
 is of the form 
$\tilde {\cal T} = {\cal T}_{\C }$ 
 for a suitable 
${\cal T} : Z_0 \oplus Z_0 \to Z_0 \oplus Z_0 $  
 if and only if 

$$ \tilde {\cal T}= C\tilde {\cal T} C $$
 or, equivalently, 
$\tilde {\cal T} $ is a {\bf cross-matrix}
 i.e., 
$\tilde {\cal T}$
 is of the form 

$$
 \tilde {\cal T} 
  =\left(
   \begin{array}{cc} \Phi & \Psi \\ 
          \overline{\Psi} & \overline{\Phi} \end{array}\right)
\mbox{ with respect to the decomposition }
 H_+ \dot+ H_- \to H_+ \dot+ H_- 
$$ 

\end{Theorem}

\vfill 

 Now, what about {\bf invariant } majorant?
 
 The theorem is:
\begin{Theorem}{ 6.}
 
 Let 
$V : \HH \to \HH $ 
 be a linear bounded invertible operator,  
 and let
$q$ 
 be an {\bf operator} majorant of 
$b(\cdot, \cdot)$.

 The following conditions are equivalent: 

 {\rm (i)}
\hfill \parbox[c]{.9\textwidth}{ $$ qV = q $$
}

 {\rm (ii)}
\hfill \parbox[c]{.9\textwidth}{ $$ qV^{-1} =q $$
}

 {\rm (iii)}
\hfill \parbox[]{.9\textwidth}{ $$ (I-R)VR = RV^{*-1}(1-R) $$
}

 {\rm (iv)}
\hfill \parbox[c]{.9\textwidth}{ $$ (I-R)V^{-1}R = RV^{*}(1-R) $$
}

 In addition, if 
$q$ 
 is a minimal majorant, and if 
$V$ 
 is an automorphism of 
$b(\cdot,\cdot)$, 
 i.e., if 
$$
 b(V\f ,V\g ) =b(\f ,\g )\,,\quad \forall \f,\g \in \HH \,, 
$$ 
 and if 
$b(\cdot,\cdot)$ 
 is symmetric
 (it means that the situation is the same as one in Example 2),
 then all conditions {\rm (i)}-{\rm (iv)} are equialent to: 

 {\rm (v)} 
\hfill \parbox[t]{.9\textwidth}{
$$
 V\left(\begin{array}{cc} I & 0 \\ K & 0 \end{array}\right)
 =\left(\begin{array}{cc} I & 0 \\ K & 0 \end{array}\right)
  V\left(\begin{array}{cc} I & 0 \\ K & 0 \end{array}\right)
$$
}

\end{Theorem}


 Let us discuss the above Theorem, especially,
 the condition (v) of this Theorem. 
 In discussing them 
 we will indicate at least three factors which have to be taken into account. 

\bigskip 
 First, we observe that the operator  
$$
  {\cal P} := \left(\begin{array}{cc} I & 0 \\ K & 0 \end{array}\right)
$$
 is a projection operator because  
${\cal P}^2 = {\cal P}$.
 Therefore the condition (v) is a condition for 
${\cal P}\HH $  
 to be a 
$V$-invariant subspace.
 We emphasize, it is true for any operator 
$V$ 
 even if 
$V$ 
 is not a Bogoliubov transformation. 
 As for the sort of the subspace, some authors refer to such subspaces 
 as the graph subspaces because one may consider 
$$
{\cal P}\HH = \{ x_+\oplus Kx_+|\,x_+ \in H_+ \} 
$$
 as the graph of the operator 
$K$. 
 In this case, 
$K$ 
 is called the {\bf angular operator } of 
${\cal P}\HH $ 
 with respect to 
$H_+$. 
%
%

\bigskip 
 The second factor is that 
$\|K\|\leq 1$.
 This inequality means, in particular, that 
 whatever 
$x \in  {\cal P}\HH $,
 the value of 
$b(x ,x ) $
 is positive: 
\begin{eqnarray*}
  b(x ,x ) &=& b(x_+\oplus Kx_+, x_+\oplus Kx_+)
            =(x_+\oplus Kx_+, \JJ( x_+\oplus Kx_+)) 
         \\&=& \|x_+\|^2 -\|Kx_+\|^2 \geq 0
\end{eqnarray*} 
 For such a sort of subspaces, there are special terms: 

 A subspace, 
$L$, 
 is called {\bf \JJ-positive} or {\bf $b$-positive} or mere {\bf positive} 
 if  
$$
  b(x ,x ) \geq 0 \,, \qquad (\forall x \in L) \,
$$ 

 Given a positive  
$L$,  
 one says `$L$ is {\bf maximal positive}', if
$L$ 
 is `set'-maximal among positive subspaces.
 In other words, a maximal positive subspace is a positive subspace   
$L$ 
 such that: whatever positive 
$L_1$ is given, 
$L_1\supset L$ 
 implies that 
$L_1=L$.

\bigskip 

 We can state now: if a quadratic-like majorant is minimal, than 
 the associated subspace is positive.  
 The property of   
$L$
 being a positive subspace, it, in itself, 
 does not implies that this subspace is of the form 
$$
 L = \{ x_+\oplus Kx_+|\,x_+ \in H_+ \}   \,, \quad \|K\|\leq 1  
$$
 but if one replaces `being a positive' by `being a maximal positive', 
 it does. 

 As the result:  
 
1) if a quadratic-like majorant is minimal, then  
 the associated subspace is maximal positive;

2) every maximal positive subspace is a subspace associated with 
 a unique minimal quadratic-like majorant.

\bigskip 

 One of the additional factors  
 which have to be taken into account in
 discussing the above Theorem 6, 
 is that a minimal majorant is regular
\footnote{ it means that $Q=R^{-1}-I$  exists and is bounded as an operator
 acting on the whole space \HH} 
 if and only if 
$\|K\| < 1$, 
 and this is exactly the case if the corresponding maximal positive subspace 
 is {\bf uniformly positive}: 
$$
 (\exists \gamma>0) (\forall x \in  {\cal P}\HH)
 \qquad b(x ,x ) \geq  \gamma b(x ,\JJ x )
$$ 
 or in other words, which are more usual for the Krein spaces theory, 
$$
 (\exists \gamma>0) (\forall x \in  {\cal P}\HH)
 \qquad < x ,x > \geq  \gamma \|x\|^2 
$$ 


\bigskip 

 The third factor which have to be taken into account  
 is that 
${\cal P}\HH $  
 is to be a 
$V$-invariant subspace.

  We will often work with the operator matrix representation of 
$V$,
$$
 V
 =\left(\begin{array}{cc} V_{11} & V_{12} \\ V_{21} & V_{22} \end{array}\right)
 \mbox{ with respect to the decomposition } 
 H_+ \oplus H_- \to H_+ \oplus H_- 
$$
 In this case the mentioned condition (v) will look like this:
$$
 \left(\begin{array}{cc} V_{11} & V_{12} \\ V_{21} & V_{22} \end{array}\right)
 \left(\begin{array}{cc} I & 0 \\ K & 0 \end{array}\right)
  =\left(\begin{array}{cc} I & 0 \\ K & 0 \end{array}\right)
 \left(\begin{array}{cc} V_{11} & V_{12} \\ V_{21} & V_{22} \end{array}\right)
 \left(\begin{array}{cc} I & 0 \\ K & 0 \end{array}\right)
$$
 One can straightforwardly verify that this condition is exactly equivalent to   
$$
   V_{21} + V_{22}K = K( V_{11} + V_{12}K )  
$$
 It is the equation which is placed among the most singular equations of the 
 Krein spaces theory 
\footnote{ if, of course, 
 one is interested in solving invariant subspaces problems } 
 and it is just the equation which we will sistematically exploit 
 when discussing Examples. 

\bigskip 

 We conclude this section with a dictionary, paralleling 
 the most explicit notions of 
 the Krein spaces theory 
 with 
 the theory of quadratic Bose Hamiltonians. 

\newpage 

\bigskip

\vspace*{\fill}

\begin{tabular}{lcl}
\hline 
\parbox[t]{0.35\textwidth}{ 
}
&  & 
\parbox[t]{0.35\textwidth}{ 
}
  \\ \medskip \\  

\parbox[t]{0.35\textwidth}{ 
 symplectic operator or form, which has the matrix
 $\Bigl(\begin{array}{cc}
          L&M\\-K&-L^T \end{array}  \Bigr)$
}
& $\longleftrightarrow $ & 
\parbox[t]{0.35\textwidth}{ 
 the quadratic Hamiltonian  
$\begin{array}{rcl} 
 h&:=&\frac{1}{2}{\sum}_{l\,m}\Bigl( 
\\&&{}
   M_{l\,m}\PP_l\PP_m
\\&&{} - L_{l\,m}(\PP_l\QQ_m + \QQ_m\PP_l) 
\\&&{} +K_{l\,m}\QQ_l\QQ_m 
                           \Bigr) 
\end{array}$ 
}
  \\ \medskip \\  
\parbox[t]{0.35\textwidth}{ 
  $J$-symmetric operator or form, which has the matrix
      $\Bigl(\begin{array}{cc}
                S & T \\ -\overline{T}  &  -\overline{S}
                   \end{array}\Bigr)$
}
& $\longleftrightarrow $ & 
\parbox[t]{0.35\textwidth}{ 
 the quadratic Hamiltonian 
 $ h={\sum}_{k,l}
 \left(s_{k\,l}a_k^*a_l-\frac{1}{2}\overline{t_{k\,l}}a_k a_l
                     -\frac{1}{2}t_{k\,l}a_l^*a_k^* \right) $
}
  \\ \medskip \\  
\parbox[t]{0.35\textwidth}{ 
 $J$-unitary operator with the matrix 
$\left(
  \begin{array}{cc} \Phi & \Psi \\ 
            \overline{\Psi} & \overline{\Phi} \end{array}\right)$
}
 & $\longleftrightarrow $ &  
\parbox[t]{0.35\textwidth}{ invertible Bogoliubov transformation 
 (quasi-free automorphism)  with the same matrix 
}
  \\ \medskip \\   
\parbox[t]{0.35\textwidth}{ 
 maximal positive subspace with the angular operator 
$K$ 
 such that 
$ K^* = \overline{K} $}
 & $\longleftrightarrow $ & 
\parbox[t]{0.35\textwidth}{ pure quadratic-like state }
  \\ \medskip \\ 
\parbox[t]{0.35\textwidth}{ 
 maximal uniformly positive subspace with the angular operator 
$K$ 
 such that 
$ K^* = \overline{K} $}
 & $\longleftrightarrow $ & 
\parbox[t]{0.35\textwidth}{ regular pure quadratic-like state, i.e.,
                            pure even quasi-free state }
  \\ \medskip \\ 
\parbox[t]{0.35\textwidth}{ 
 maximal positive invariant subspace with the angular operator 
$K$ 
 such that 
$ K^* = \overline{K} $
 }
 & $\longleftrightarrow $ & 
\parbox[t]{0.35\textwidth}{ pure quadratic-like invariant state }
  \\ \medskip \\ 
\hline 
\end{tabular}

\bigskip 

 Now then, it is time to Examples.  

 
\begin{Thm}{COMMENT}{ } \rm 

 About linear canonical transformations and hamiltonians, see e.g.,    
 [W1,2,3] for $dim < \infty$ and e.g., [Ber], [BR2], [RS2] for 
 the quantum case.
 Theorem 5 see in [Ber], see also [DK], [K].
 The standard point is concentrated on the questions 
 ``how diagonalize a given hamiltonian or automophism ?'' and 
 ``does there exist a {\bf regular} invariant state ?''
 We interested in any invariant states 
 no matter whether they are regular or not 
 and 
 any hamiltonians 
 no matter whether they are diagonalizable or not. 

 For terms `angular operator', `positive subspace', `maximal positive subspace'
 and for other details of the Krein spaces theory, 
 see,  e.g., [Bog], [DR].  

 The approach in this section  
 is based on [Ch1], [Ch2], [Ch3] and the summarizing [Ch4].

\end{Thm}

\newpage

\newpage 
\section{ Examples }
\subsection{ Example 1. Oscillator }

 In terms of 
$\PP, \QQ$ , 
 the Hamiltonian is written as 
$$
 h:=\frac{1}{2}\PP^2 +\frac{1}{2}\Omega_0^2\QQ^2
$$ 
 Then  
$$
 i[h,F(x_p\oplus x_q)]=F(x_{p}'\oplus x_{q}')
$$
 where 
$$
 {x_{p}' \choose x_{q}'} 
     = \Bigl(\begin{array}{cc}
             L&M\\-K&-L^T \end{array}  \Bigr) {x_{p}\choose x_{q}} 
$$

$$
  \Bigl(\begin{array}{cc}     L&M\\-K&-L^T \end{array}  \Bigr) 
   = \Bigl(\begin{array}{cc}  0&1\\-\Omega_0^2&0 \end{array}  \Bigr) 
$$
$$
   V_t 
   := e^{ t \Bigl(\begin{array}{cc}  0&1\\-\Omega_0^2&0 \end{array}  \Bigr) }
   = \Bigl(\begin{array}{cc} 
  cos(\Omega_0 t)&\Omega_0^{-1}sin(\Omega_0 t)\\
   -\Omega_0 sin(\Omega_0 t)&cos(\Omega_0 t) 
 \end{array}  \Bigr) 
$$

 In terms of 
$a^*, a$ , 
 the Hamiltonian is rewritten as 

\begin{eqnarray*}
 h &:= &\frac{1}{2}\PP^2 +\frac{1}{2}\Omega_0^2\QQ^2 \\
   & = &\frac{1}{2}(\frac{i}{\sqrt 2}(a^* - a))^2 
       + \frac{1}{2}\Omega_0^2(\frac{1}{\sqrt 2}(a^* + a))^2 \\
   & = &\frac{1+\Omega_0^2}{2} a^* a
        -\frac{1-\Omega_0^2}{4} a^{*2} -\frac{1-\Omega_0^2}{4} a^2 
        + const  \\ 
\end{eqnarray*}

 Formal calculations show that 
$$
 [h,A(u^+\oplus u^-)]=A({u^+}'\oplus {u^-}')\,,
$$
 where 
${u^+}'\oplus {u^-}'$ 
 is defined by
$$
 {{u^+}'\choose {u^-}'}
    =\Bigl(\begin{array}{cc}
                S & T \\ -\overline{T}  &  -\overline{S}
                   \end{array}\Bigr)
 {u^+\choose u^-}
$$
$$
 \Bigl(\begin{array}{cc}
    S & T \\ -\overline{T}  &  -\overline{S}
       \end{array}\Bigr)
 = \Bigl(\begin{array}{cc}
      \frac{1+\Omega_0^2}{2} & \frac{1-\Omega_0^2}{2} \\ 
     -\frac{1-\Omega_0^2}{2} &  -\frac{1+\Omega_0^2}{2}
        \end{array}\Bigr)
$$
$$
 -\frac{1-\Omega_0^2}{2}  -\frac{1+\Omega_0^2}{2} K
   =   K \frac{1+\Omega_0^2}{2} + K\frac{1-\Omega_0^2}{2}K 
$$
$$
 -(1-\Omega_0^2) = 2K (1+\Omega_0^2) + (1-\Omega_0^2)K^2 
$$
 Recall that 
$\|K\| \leq 1$ 
 and assume 
$\Omega_0 > 0$ . 
 Then 
$$
     K =-\frac{1-\Omega_0}{1+\Omega_0}
$$
 is a {\bf unique } solution.  
 As for the corresponding
$R, Q, q, \omega $,
 we have in terms of 
$\PP, \QQ$, 

$$ 
 R 
  =\frac{1}{4}
\left(
\begin{array}{cc}
  2 - K - K^*  &   iK - iK^* \\ 
  iK - iK^*    &   2 + K + K^* 
\end{array}
\right)
 =
\left(
\begin{array}{cc}
  \frac{1}{1+\Omega_0}  &          0 \\ 
             0          &   \frac{\Omega_0}{1+\Omega_0} 
\end{array}
\right)
$$ 
$$
  Q 
  = 
\left(
\begin{array}{cc}
 \Omega_0  &           0 \\ 
    0      &   \frac{1}{\Omega_0} 
\end{array}
\right)
$$

$$
  \omega e^{\displaystyle ix_{p}\PP + ix_{q}\QQ }
 =  e^{-\displaystyle q(x_{p}\oplus x_{q})/4 }
 =  e^{-\displaystyle (\Omega_0x_{p}^2 + \frac{1}{\Omega_0}x_{q}^2)/4 }
$$

 A few detailes of
 asymptotic behaviour of 
$\alpha_t := \alpha_{V_t}$ 
 are the folllowing:
 Consider the standard Fock state, i.e., 
 the state, 
$\omega_F$, 
 defined by  
$$
  \omega_F e^{\displaystyle ix_{p}\PP + ix_{q}\QQ }
 =  e^{-|x_{p}\oplus x_{q}|^2/4 }
 =  e^{-\displaystyle (x_{p}^2 + x_{q}^2)/4 }
$$

 Then 

\begin{eqnarray*}
\makebox[4ex][l]{
 $\omega_F \alpha_t e^{\displaystyle ix_{p}\PP + ix_{q}\QQ } $
}
\\ &=&  e^{-|V_t(x_{p}\oplus x_{q})|^2/4 } 
\\ &=&  e^{-\displaystyle (
  (cos(\Omega_0 t)x_{p}+\Omega_0^{-1}sin(\Omega_0 t)x_{q})^2
  +( -\Omega_0 sin(\Omega_0 t)x_{p}+cos(\Omega_0 t)x_{q})^2 
 )/4 }
\end{eqnarray*}
 We see, this quantity has no usual limit, 
 neither as 
$t \to +\infty$ 
 nor as 
$t \to -\infty$ ,
 whenever 
$\Omega_0 \not= \pm 1$. 

\newpage 
\subsection{ Example 2. Free Evolution on Line }

 In terms of 
$\PP, \QQ$ , 
 the Hamiltonian is written as 
$$
 h:=\frac{1}{2}\PP^2  
$$ 
 Then  
$$
 i[h,F(x_p\oplus x_q)]=F(x_{p}'\oplus x_{q}')
$$
 where 
$$
 {x_{p}' \choose x_{q}'} 
     = \Bigl(\begin{array}{cc}
             L&M\\-K&-L^T \end{array}  \Bigr) {x_{p}\choose x_{q}} 
$$

$$
  \Bigl(\begin{array}{cc}     L&M\\-K&-L^T \end{array}  \Bigr) 
   = \Bigl(\begin{array}{cc}  0&1\\0&0 \end{array}  \Bigr) 
$$
$$
 V_t 
  := e^{ t \Bigl(\begin{array}{cc}  0&1\\0&0 \end{array}  \Bigr) }
  = \Bigl(\begin{array}{cc} 
      1 & t\\
      0 & 1 
  \end{array}  \Bigr) 
$$

 In terms of 
$a^*, a$ , 
 the Hamiltonian is rewritten as 

\begin{eqnarray*}
 h &:= &\frac{1}{2}\PP^2 \\
   & = &\frac{1}{2}(\frac{i}{\sqrt 2}(a^* - a))^2 \\
   & = &\frac{1}{2} a^* a
        -\frac{1}{4} a^{*2} -\frac{1}{4} a^2 
        + const  \\ 
\end{eqnarray*}
 Formal calculations show that 
$$
 [h,A(u^+\oplus u^-)]=A({u^+}'\oplus {u^-}')\,,
$$
 where 
${u^+}'\oplus {u^-}'$ 
 is defined by
$$
 {{u^+}'\choose {u^-}'}
    =\Bigl(\begin{array}{cc}
                S & T \\ -\overline{T}  &  -\overline{S}
                   \end{array}\Bigr)
 {u^+\choose u^-}
$$
$$
 \Bigl(\begin{array}{cc}
    S & T \\ -\overline{T}  &  -\overline{S}
       \end{array}\Bigr)
 = \Bigl(\begin{array}{cc}
      \frac{1}{2} & \frac{1}{2} \\ 
     -\frac{1}{2} &  -\frac{1}{2}
        \end{array}\Bigr)
$$
$$
 -\frac{1}{2}  -\frac{1}{2} K
   =   K \frac{1}{2} + K\frac{1}{2}K 
$$
$$
 -1 = 2K  + K^2 
$$
 Then 
$$
     K = -1 
$$
 is a {\bf unique } solution.  
 As for the corresponding
$R, Q, q, \omega $,
 we have in terms of 
$\PP, \QQ$, 

$$ 
 R 
  =\frac{1}{4}
\left(
\begin{array}{cc}
  2 - K - K^*  &   iK - iK^* \\ 
  iK - iK^*    &   2 + K + K^* 
\end{array}
\right)
 =
\left(
\begin{array}{cc}
 1 & 0 \\ 
 0 & 0 
\end{array}
\right)
$$ 

$$
  \omega e^{\displaystyle ix_{p}\PP + ix_{q}\QQ }
 =  e^{-\displaystyle q(x_{p}\oplus x_{q})/4 }
 =  e^{-\displaystyle \infty\cdot x_{q}^2/4 }
 =\left\{\begin{array}{cl}
         1,& \mbox{ if } x_{q} =0 \\
         0,& \mbox{ if } x_{q} \not=0 \\
         \end{array}\right. 
$$

 A few detailes of
 asymptotic behaviour of 
$\alpha_t := \alpha_{V_t}$ 
 are the folllowing:
 Consider the standard Fock state, i.e., 
 the state, 
$\omega_F$, 
 defined by  
$$
  \omega_F e^{\displaystyle ix_{p}\PP + ix_{q}\QQ }
 =  e^{-|x_{p}\oplus x_{q}|^2/4 }
 =  e^{-\displaystyle (x_{p}^2 + x_{q}^2)/4 }
$$

 Then 
\begin{eqnarray*}
  \omega_F \alpha_t e^{\displaystyle ix_{p}\PP + ix_{q}\QQ }
   &=&  e^{-|V_t(x_{p}\oplus x_{q})|^2/4 } 
\\ &=&  e^{-\displaystyle (
  (x_{p}+ tx_{q})^2
  +x_{q}^2 
 )/4 } 
\end{eqnarray*}

 We see, this quantity has a limit
 as 
$t \to +\infty$ 
 and as 
$t \to -\infty$ 
 as well
 and these limits are equal: 

\begin{eqnarray*}
  \lim_{t\to \pm\infty}
  \omega_F \alpha_t e^{\displaystyle ix_{p}\PP + ix_{q}\QQ }
  &=&  \lim_{t\to \pm\infty}
  e^{-\displaystyle (
  (x_{p}+ tx_{q})^2
  +x_{q}^2 
  )/4 } \\
 &=& 
  \left\{\begin{array}{cl}
  e^{-x_{p}^2/4},& \mbox{ if } x_{q} =0 \\
         0,        & \mbox{ if } x_{q} \not=0 \\
         \end{array}\right. 
\end{eqnarray*}
 Notice,  
$$
  \lim_{t\to \pm\infty}
  \omega_F \alpha_t e^{\displaystyle ix_{p}\PP + ix_{q}\QQ }
$$
 is not a charactiristic functional of a {\bf pure} state.

\newpage 
\subsection{ Example 3.   $h:=\frac{1}{2}(\PP \QQ$ + \QQ \PP )}

 In terms of 
$\PP, \QQ$ , 
 the Hamiltonian is written as 
$$
 h:=\frac{1}{2}(\PP \QQ + \QQ \PP ) 
$$ 
 Then  
$$
 i[h,F(x_p\oplus x_q)]=F(x_{p}'\oplus x_{q}')
$$
 where 
$$
 {x_{p}' \choose x_{q}'} 
     = \Bigl(\begin{array}{cc}
             L&M\\-K&-L^T \end{array}  \Bigr) {x_{p}\choose x_{q}} 
$$

$$
  \Bigl(\begin{array}{cc}     L&M\\-K&-L^T \end{array}  \Bigr) 
   = \Bigl(\begin{array}{cc}  -1&0\\0&1 \end{array}  \Bigr) 
$$
$$
   V_t 
   := e^{ t \Bigl(\begin{array}{cc}  -1&0\\0&1 \end{array}  \Bigr) }
   = \Bigl(\begin{array}{cc} 
  e^{-t} & 0\\
      0& e^{t} 
 \end{array}  \Bigr) 
$$

 In terms of 
$a^*, a$ , 
 the Hamiltonian is rewritten as 

\begin{eqnarray*}
 h  &:= &\frac{1}{2}(\PP \QQ + \QQ \PP )
 \\ & = &\frac{1}{2}
        \left(\frac{i}{\sqrt 2}(a^* - a)\cdot \frac{1}{\sqrt 2}(a^* + a) 
         +\frac{i}{\sqrt 2}(a^* + a)\cdot \frac{1}{\sqrt 2}(a^* - a)\right) 
 \\ & = & \frac{i}{2} a^{*2} -\frac{i}{2} a^2 
 \\ 
\end{eqnarray*}

 Formal calculations show that 
$$
 [h,A(u^+\oplus u^-)]=A({u^+}'\oplus {u^-}')\,,
$$
 where 
${u^+}'\oplus {u^-}'$ 
 is defined by
$$
 {{u^+}'\choose {u^-}'}
    =\Bigl(\begin{array}{cc}
                S & T \\ -\overline{T}  &  -\overline{S}
                   \end{array}\Bigr)
 {u^+\choose u^-}
$$
$$
 \Bigl(\begin{array}{cc}
    S & T \\ -\overline{T}  &  -\overline{S}
       \end{array}\Bigr)
 = \Bigl(\begin{array}{cc}
      0  & -i \\ 
     -i  &  0 
        \end{array}\Bigr)
$$
$$
  -i =  K\cdot (-i)\cdot K 
$$
 Then there are {\bf two (!)} solutions: 
$$
     K = K_{+1} =  1
$$
$$
     K = K_{-1} = -1
$$

\newpage 

 As for the corresponding
$R_{+1}, Q_{+1}, q_{+1}, \omega_{+1} $,
 and 
$R_{-1}, Q_{-1}, q_{-1}, \omega_{-1} $,
 we have in terms of 
$\PP, \QQ$, 

$$ 
 R_{+1} 
  =\frac{1}{4}
\left(
\begin{array}{cc}
  2 - K_{+1} - K_{+1}^*  &   iK_{+1} - iK_{+1}^* \\ 
  iK_{+1} - iK_{+1}^*    &   2 + K_{+1} + K_{+1}^* 
\end{array}
\right)
 =
\left(
\begin{array}{cc}
 0 & 0 \\ 
 0 & 1 
\end{array}
\right)
$$ 

$$
  \omega_{+1} e^{\displaystyle ix_{p}\PP + ix_{q}\QQ }
 =  e^{-\displaystyle q_{+1}(x_{p}\oplus x_{q})/4 }
 =  e^{-\displaystyle \infty\cdot x_{p}^2/4 }
 =\left\{\begin{array}{cl}
         1,& \mbox{ if } x_{p} =0 \\
         0,& \mbox{ if } x_{p} \not=0 \\
         \end{array}\right. 
$$

$$ 
 R_{-1} 
  =\frac{1}{4}
\left(
\begin{array}{cc}
  2 - K_{-1} - K_{-1}^*  &   iK_{-1} - iK_{-1}^* \\ 
  iK_{-1} - iK_{-1}^*    &   2 + K_{-1} + K_{-1}^* 
\end{array}
\right)
 =
\left(
\begin{array}{cc}
 1 & 0 \\ 
 0 & 0 
\end{array}
\right)
$$ 

$$
  \omega_{-1} e^{\displaystyle ix_{p}\PP + ix_{q}\QQ }
 =  e^{-\displaystyle q_{-1}(x_{p}\oplus x_{q})/4 }
 =  e^{-\displaystyle \infty\cdot x_{q}^2/4 }
 =\left\{\begin{array}{cl}
         1,& \mbox{ if } x_{q} =0 \\
         0,& \mbox{ if } x_{q} \not=0 \\
         \end{array}\right. 
$$

 If we confer these expressions with that in Example 1,
$$
  \omega e^{\displaystyle ix_{p}\PP + ix_{q}\QQ }
 =  e^{-\displaystyle q(x_{p}\oplus x_{q})/4 }
 =  e^{-\displaystyle (\Omega_0x_{p}^2 + \frac{1}{\Omega_0}x_{q}^2)/4 }
$$
 we can infer that 
$\omega_{+1}$ 
 is an approximation of the ground state of an oscillator 
 with the extremely high frequency 
$\Omega_0$
 whereas 
$\omega_{-1}$ 
 is an approximation of the ground state of an oscillator 
 with the extremely low frequency 
$\Omega_0$.

 Finally, we have 
\begin{eqnarray*}
  \lim_{t\to +\infty}
  \omega_F \alpha_t e^{\displaystyle ix_{p}\PP + ix_{q}\QQ }
 &=&  \lim_{t\to +\infty}
  e^{-\displaystyle (
  e^{-2t}x_{p}^2 + e^{2t}x_{q}^2
  )/4 } \\
 &=& 
  \left\{\begin{array}{cl}
         1 ,& \mbox{ if } x_{q} =0 \\
         0,        & \mbox{ if } x_{q} \not=0 \\
         \end{array}\right. \\ 
 &=& \omega_{-1}e^{\displaystyle ix_{p}\PP + ix_{q}\QQ }
\end{eqnarray*}

\begin{eqnarray*}
  \lim_{t\to -\infty}
  \omega_F \alpha_t e^{\displaystyle ix_{p}\PP + ix_{q}\QQ }
 &=&  \lim_{t\to -\infty}
  e^{-\displaystyle (
  e^{-2t}x_{p}^2 + e^{2t}x_{q}^2
  )/4 } \\
 &=& 
  \left\{\begin{array}{cl}
         1 ,& \mbox{ if } x_{p} =0 \\
         0,        & \mbox{ if } x_{p} \not=0 \\
         \end{array}\right. \\ 
 &=& \omega_{+1}e^{\displaystyle ix_{p}\PP + ix_{q}\QQ }
\end{eqnarray*}
 This situation is typical. Whatever {\bf regular} quadratic state
$\omega $, 
$$
  \omega e^{\displaystyle ix_{p}\PP + ix_{q}\QQ }
 = e^{-\displaystyle (q_{11}x_{p}^2 + 2q_{22}x_{p}x_{q} + q_{22}x_{q}^2)/4 } 
 \,, 
$$
 we have choosen, the result is: 

\begin{eqnarray*}
  \lim_{t\to +\infty}
  \omega \alpha_t e^{\displaystyle ix_{p}\PP + ix_{q}\QQ }
 &=& 
  \left\{\begin{array}{cl}
         1 ,& \mbox{ if } x_{q} =0 \\
         0,        & \mbox{ if } x_{q} \not=0 \\
         \end{array}\right. \\ 
 &=& \omega_{-1}e^{\displaystyle ix_{p}\PP + ix_{q}\QQ }
\end{eqnarray*}

\begin{eqnarray*}
  \lim_{t\to -\infty}
  \omega \alpha_t e^{\displaystyle ix_{p}\PP + ix_{q}\QQ }
 &=& 
  \left\{\begin{array}{cl}
         1 ,& \mbox{ if } x_{p} =0 \\
         0,        & \mbox{ if } x_{p} \not=0 \\
         \end{array}\right. \\ 
 &=& \omega_{+1}e^{\displaystyle ix_{p}\PP + ix_{q}\QQ }
\end{eqnarray*}

\newpage 
\subsection{ Example 4. Repulsive Oscillator }

 In terms of 
$\PP, \QQ$ , 
 the Hamiltonian is written as 
$$
 h:=\frac{1}{2}\PP^2 -\frac{1}{2}\Omega_0^2\QQ^2
$$ 
 Then  
$$
 i[h,F(x_p\oplus x_q)]=F(x_{p}'\oplus x_{q}')
$$
 where 
$$
 {x_{p}' \choose x_{q}'} 
     = \Bigl(\begin{array}{cc}
             L&M\\-K&-L^T \end{array}  \Bigr) {x_{p}\choose x_{q}} 
$$

$$
  \Bigl(\begin{array}{cc}     L&M \\-K&-L^T \end{array}  \Bigr) 
   = \Bigl(\begin{array}{cc}  0&1 \\ \Omega_0^2&0 \end{array}  \Bigr) 
$$
$$
 V_t 
  := e^{ t \Bigl(\begin{array}{cc}  0&1\\-\Omega_0^2&0 \end{array}  \Bigr) }
  = \Bigl(\begin{array}{cc} 
  cos(i\Omega_0 t)&(i\Omega_0)^{-1}sin(i\Omega_0 t)\\
  -i\Omega_0 sin(i\Omega_0 t)&cos(i\Omega_0 t) 
 \end{array}  \Bigr) 
$$
$$
 V_t 
  := e^{ t \Bigl(\begin{array}{cc}  0&1\\-\Omega_0^2&0 \end{array}  \Bigr) }
  = \frac{1}{2}\Bigl(\begin{array}{cc} 
  e^{\Omega_0 t}+e^{-\Omega_0 t}&
  \Omega_0^{-1}( e^{\Omega_0 t}-e^{-\Omega_0 t} )\\
  \Omega_0( e^{\Omega_0 t}-e^{-\Omega_0 t} ) &
  e^{\Omega_0 t}+e^{-\Omega_0 t} 
 \end{array}  \Bigr) 
$$

 In terms of 
$a^*, a$ , 
 the Hamiltonian is rewritten as 

\begin{eqnarray*}
 h &:= &\frac{1}{2}\PP^2 -\frac{1}{2}\Omega_0^2\QQ^2 \\
   & = &\frac{1}{2}(\frac{i}{\sqrt 2}(a^* - a))^2 
       - \frac{1}{2}\Omega_0^2(\frac{1}{\sqrt 2}(a^* + a))^2 \\
   & = &\frac{1-\Omega_0^2}{2} a^* a
        -\frac{1+\Omega_0^2}{4} a^{*2} -\frac{1+\Omega_0^2}{4} a^2 
        + const  \\ 
\end{eqnarray*}

 Formal calculations show that 
$$
 [h,A(u^+\oplus u^-)]=A({u^+}'\oplus {u^-}')\,,
$$
 where 
${u^+}'\oplus {u^-}'$ 
 is defined by
$$
 {{u^+}'\choose {u^-}'}
    =\Bigl(\begin{array}{cc}
                S & T \\ -\overline{T}  &  -\overline{S}
                   \end{array}\Bigr)
 {u^+\choose u^-}
$$
$$
 \Bigl(\begin{array}{cc}
    S & T \\ -\overline{T}  &  -\overline{S}
       \end{array}\Bigr)
 = \Bigl(\begin{array}{cc}
      \frac{1-\Omega_0^2}{2} & \frac{1+\Omega_0^2}{2} \\ 
     -\frac{1+\Omega_0^2}{2} &  -\frac{1-\Omega_0^2}{2}
        \end{array}\Bigr)
$$
$$
 -\frac{1+\Omega_0^2}{2}  -\frac{1-\Omega_0^2}{2} K
   =   K \frac{1-\Omega_0^2}{2} + K\frac{1+\Omega_0^2}{2}K 
$$
$$
 -(1+\Omega_0^2) = 2K (1-\Omega_0^2) + (1+\Omega_0^2)K^2 
$$
 Recall that 
$\|K\| \leq 1$ 
 and assume 
$\Omega_0 > 0$ . 
 Then there are {\bf two (!)} solutions: 
$$
     K = K_{+1} = -\frac{1-i\Omega_0}{1+i\Omega_0}
$$
$$
     K = K_{-1} = -\frac{1+i\Omega_0}{1-i\Omega_0}
$$

\newpage 

 As for the corresponding
$R_{+1}, Q_{+1}, q_{+1}, \omega_{+1} $,
 and 
$R_{-1}, Q_{-1}, q_{-1}, \omega_{-1} $,
 we have in terms of 
$\PP, \QQ$, 

$$ 
 R_{+1} 
  =\frac{1}{4}
\left(
\begin{array}{cc}
  2 - K_{+1} - K_{+1}^*  &   iK_{+1} - iK_{+1}^* \\ 
  iK_{+1} - iK_{+1}^*    &   2 + K_{+1} + K_{+1}^* 
\end{array}
\right)
 = \frac{1}{1+\Omega_0^2}
\left(
\begin{array}{cc}
     1      &  \Omega_0 \\ 
 \Omega_0   &  \Omega_0^2 
\end{array}
\right)
$$ 

\begin{eqnarray*}
  \omega_{+1} e^{\displaystyle ix_{p}\PP + ix_{q}\QQ }
 &=&  e^{-\displaystyle q_{+1}(x_{p}\oplus x_{q})/4 }
 \\&=&  e^{-\displaystyle \infty\cdot (-\Omega_0x_{p}+x_{q})^2/4 }
    = \left\{\begin{array}{cl}
         1,& \mbox{ if } -\Omega_0x_{p}+x_{q} =0 \\
         0,& \mbox{ if } -\Omega_0x_{p}+x_{q} \not=0 \\
         \end{array}\right. 
\end{eqnarray*}

$$ 
 R_{-1} 
  =\frac{1}{4}
\left(
\begin{array}{cc}
  2 - K_{-1} - K_{-1}^*  &   iK_{-1} - iK_{-1}^* \\ 
  iK_{-1} - iK_{-1}^*    &   2 + K_{-1} + K_{-1}^* 
\end{array}
\right)
 = \frac{1}{1+\Omega_0^2}
\left(
\begin{array}{cc}
     1      &  -\Omega_0 \\ 
 -\Omega_0  &   \Omega_0^2 
\end{array}
\right)
$$ 

\begin{eqnarray*}
  \omega_{-1} e^{\displaystyle ix_{p}\PP + ix_{q}\QQ }
 &=&  e^{-\displaystyle q_{-1}(x_{p}\oplus x_{q})/4 }
 \\&=&  e^{-\displaystyle \infty\cdot (\Omega_0x_{p}+x_{q})^2/4 }
    = \left\{\begin{array}{cl}
         1,& \mbox{ if } \Omega_0x_{p}+x_{q} =0 \\
         0,& \mbox{ if } \Omega_0x_{p}+x_{q} \not=0 \\
         \end{array}\right. 
\end{eqnarray*}

 Finally, one can verify that  
\begin{eqnarray*}
  \lim_{t\to +\infty}
  \omega_F \alpha_t e^{\displaystyle ix_{p}\PP + ix_{q}\QQ }
 &=& \omega_{-1}e^{\displaystyle ix_{p}\PP + ix_{q}\QQ }
\end{eqnarray*}

\begin{eqnarray*}
  \lim_{t\to -\infty}
  \omega_F \alpha_t e^{\displaystyle ix_{p}\PP + ix_{q}\QQ }
 &=& \omega_{+1}e^{\displaystyle ix_{p}\PP + ix_{q}\QQ }
\end{eqnarray*}
 This situation is typical as in the previous Example. 
 Whatever {\bf regular} quadratic state
$\omega $, 
$$
  \omega e^{\displaystyle ix_{p}\PP + ix_{q}\QQ }
 = e^{-\displaystyle (q_{11}x_{p}^2 + 2q_{22}x_{p}x_{q} + q_{22}x_{q}^2)/4 } 
 \,, 
$$
 we have choosen, the result is: 

\begin{eqnarray*}
  \lim_{t\to +\infty}
  \omega \alpha_t e^{\displaystyle ix_{p}\PP + ix_{q}\QQ }
 &=& \omega_{-1}e^{\displaystyle ix_{p}\PP + ix_{q}\QQ }
\end{eqnarray*}

\begin{eqnarray*}
  \lim_{t\to -\infty}
  \omega \alpha_t e^{\displaystyle ix_{p}\PP + ix_{q}\QQ }
 &=& \omega_{+1}e^{\displaystyle ix_{p}\PP + ix_{q}\QQ }
\end{eqnarray*}

\newpage 
\subsection{ Example 5.    }


 consider the approximating (Bogoliubov) 
 Hamiltonian 
\begin{eqnarray*} 
 H_B' 
 &=& \int dp\,\left\{\omega(p)a^*(p)a(p) + 
 \frac{1}{2}\Delta_B(p)\left[a(p)^*a(-p)^* + a(-p)a(p)\right]
             \right\}  \\ 
\end{eqnarray*}

 Formal calculations show that 
$$
 [H_B',A(u^+\oplus u^-)]=A({u^+}'\oplus {u^-}')\,,
$$
 where 
${u^+}'\oplus {u^-}'$ 
 is defined by
$$
 {{u^+}'\choose {u^-}'}
    =\Bigl(\begin{array}{cc}
                S & T \\ -\overline{T}  &  -\overline{S}
                   \end{array}\Bigr)
 {u^+\choose u^-}
$$
$$
 \Bigl(\begin{array}{cc}
    S & T \\ -\overline{T}  &  -\overline{S}
       \end{array}\Bigr)(p,p')
 = \Bigl(\begin{array}{cc}
      \omega(p)\delta(p-p')   & -\Delta_B(p)\delta(p+p') \\ 
      \Delta_B(p)\delta(p+p') & -\omega(p)\delta(p-p') 
        \end{array}\Bigr)
$$
$$
 \Bigl(\begin{array}{cc}
    S & T \\ -\overline{T}  &  -\overline{S}
       \end{array}\Bigr)
 = \Bigl(\begin{array}{cc}
      \hat \omega     & -\hat\Delta_B J_0 \\ 
      J_0\hat\Delta_B & -\hat \omega 
        \end{array}\Bigr)
$$
$$
  J_0\hat\Delta_B  -\hat \omega K 
  = K \hat \omega - K \hat\Delta_B J_0 K 
$$
$$
  -\hat\Delta_B  +J_0\hat \omega J_0 K 
  = -J_0 K \hat \omega + J_0K \hat\Delta_B J_0 K 
$$
 If we take into account that 
$\omega(-p)=\omega(p)$,  
 i.e., 
$J_0\hat \omega J_0 = \hat \omega $  
 and if we restrict ourselves to the case where 
$J_0K$ 
 commutes with the multiplications by functions, i.e., if
$K$ 
 is of the form 
$$
  K(p,p')
  = \delta(p+p')k_0(p)
$$
 for a function 
$k_0$, 
 then we obtain:
$$
 -\Delta_B(p) 
  = -2\omega(p) k_0(p) 
       + \Delta_B(p) k_0(p)^2\,,\, |k_0(p)|\leq 1\,. 
$$
 Thus 
$$
 k_0(p) 
  = \left\{ \begin{array}{cl}
  0,& \mbox{ if $p$ is such that }\\
    & \Delta_B(p) = 0, \omega(p) \not= 0\\
  arbitrary ,& \mbox{ if $p$ is such that }\\
    &  \omega(p) = 0, \Delta(p) = 0 \\
\displaystyle 
\frac{\omega(p) 
  - sgn(\omega(p))\sqrt{ -\Delta_B(p)^2 + \omega(p)^2}}{\Delta_B(p)},
 & \mbox{ if $p$ is such that }\\
 & -\Delta_B(p)^2 + \omega(p)^2 \geq 0, \Delta(p) \not= 0  \\
\displaystyle 
\frac{\omega(p) 
       -i\epsilon(p) \sqrt{ \Delta_B(p)^2 - \omega(p)^2}}{\Delta_B(p)},
 & \mbox{ if $p$ is such that }\\ 
\mbox{ where } \epsilon(p)^2 = 1  
 &-\Delta_B(p)^2 + \omega(p)^2 \leq 0, \Delta(p) \not= 0 \\
                       \end{array}\right.
$$
 These relationships can be transformed as follows: 
$$
 k_0(p) 
  = \left\{ \begin{array}{cl}
  arbitrary ,& \mbox{ if $p$ is such that }\\
    &  \omega(p) = 0, \Delta(p) = 0 \\
\displaystyle 
\frac{\Delta_B(p)}{\omega(p) 
                   + sgn(\omega(p))\sqrt{ -\Delta_B(p)^2 + \omega(p)^2}},
 & \mbox{ if $p$ is such that }\\
 & -\Delta_B(p)^2 + \omega(p)^2 \geq 0 \\
\displaystyle 
\frac{\Delta_B(p)}{\omega(p) 
                    +i\epsilon(p) \sqrt{ \Delta_B(p)^2 - \omega(p)^2}},
 & \mbox{ if $p$ is such that }\\ 
\mbox{ where } \epsilon(p)^2 = 1  &-\Delta_B(p)^2 + \omega(p)^2 \leq 0 \\
                       \end{array}\right.
$$

 In particular, if there are infinitely many 
$p$ 
 such that 
$-\Delta_B(p)^2 + \omega(p)^2 \leq 0$, 
 then there are infinitely many 
 invariant pure quadratic-like states.

                             \newpage

\section{Appendix A: The Formal Calculations of $e^{A+B}$}

 Assume
$$
 [[A,B],B]= 0\,,\,  [[A,B],A]= 0\,. 
$$

 Now let us calculate 
$$
 U:=e^{t(A+B)}.
$$
 The definition of 
$$
 U
$$
 formally implies
$$
 dU/dt = (A + B)U\,,\, U(0) = I\,.
$$
 Let
$$
  U:= e^{tA}V
$$
 Hence, 
$$
  dV/dt = e^{-tA}Be^{tA}V = (B-t[A,B])V \,,\, V(0)=I \,.
$$
 Let
$$
  V:= e^{tB}C\,.
$$
 Then,
$$
 dC/dt = -e^{-tB}t[A,B]e^{tB} =-t[A,B]C  \,,\, C(0)=I \,.
$$
 Hence 
$$
  C=e^{-\frac{t^2}{2}[A,B]} \,.
$$
\subsection*{ the Result is:}

$$
  e^{t(A+B)} =e^{tA}e^{tB}e^{-\frac{t^2}{2}[A,B]}
$$

$$
  e^{i(A+B)} =e^{iA}e^{iB}e^{\frac{i}{2}[A,B]}
$$
$$
  e^{i(A+B)} =e^{iA}e^{iB}e^{\frac{1}{2}[A,B]}
$$

$$
  e^{A}e^{B}=e^{\frac{1}{2}[A,B]}e^{A+B} 
$$
$$
  e^{iA}e^{iB}=e^{-\frac{1}{2}[A,B]}e^{i(A+B)}
$$

\begin{Remark}{ .}

 The usual form of the CCRs is motivated by the Schr\"odinger representation
 of the position and momentum operators:
$$
  \QQ =\hat{x} \quad,\quad  \PP =\frac{\hbar}{i}\frac{\partial}{\partial x} \,.
$$
 Hence, 
$$
  \frac{i}{\hbar}[\PP ,\QQ ] = 1 \,,\, [\QQ ,\PP ] = i\hbar \,,\,
 [\frac{1}{2}{\PP }^2,\QQ ] =
 \frac{1}{2}\cdot 2\cdot \PP \cdot \frac{\hbar}{i}= \frac{\hbar}{i}\PP \,.
$$

\end{Remark}

\newpage

\bibliographystyle{unsrt}

\end{document}